# A Quantum Copy-Protection Scheme with Authentication

Laszlo Gyongyosi[*], Sandor Imre

*Department of Telecommunications, Budapest University of Technology and Economics*

Magyar tudosok krt. 2., Budapest, H-1117, Hungary

[*]*gyongyosi@hit.bme.hu*

(Dated: 2009)

**Abstract:** We propose a quantum copy-protection system which protects classical information in the form of non-orthogonal quantum states. The decryption of the stored information is not possible in the classical representation and the decryption mechanism of data qubits is realized by secret unitary rotations. We define an authentication method for the proposed copy-protection scheme and analyse the success probabilities of the authentication process. A possible experimental realization of the scheme is also presented.

## 1  Introduction

In classical computer science, any data that can be read can be copied an unlimited number of times. There is no theoretical limit on the copying of classical data within classical communication. The best solution we can expect is that the media that carry these data should not be duplicable. On the other hand, the rise of quantum information processing reveals that with quantum information a higher level of security can be expected. The entertainment and classical computer industry use every technical and legal means possible to prevail against pirates and the main purpose is to implement widespread copy prevention of digital files. However, cloning a quantum state is an impossible task and this is called the no-cloning theorem (Wootters and Zurek, 1982). The no-cloning theorem was first shown by Wooters and Zurek and this theory is closely concerned with other quantum information processing such as state estimation and quantum cloning. In the quantum world, the quantum states cannot be copied perfectly, according to the no-cloning theorem (Nielsen, Chuang, 2000). The proposed quantum-based data protection system uses the fundamental difference between classical and quantum information (Imre and Balazs, 2005). The classical data are stored in the secret rotation angles of non-orthogonal quantum states. We show that secret quantum decoding operations can be implemented with some associated error, which decreases exponentially with the number of quantum states of the angle state.

## 2  Quantum encryption

In the classical world, copy-protecting is trivially impossible. In this paper we show a new quantum-based data protection system, which stores the classical data inside the encoded quantum states on the data medium. In the proposed quantum-based data protection system, Alice can always read the *classical data*, while no one but the original *Issuer* knows the original quantum states that carry these classical data. Even though an unauthorized user can try to copy some other quantum states storing the same classical data, these quantum states can be distinguished from the original ones. The classical data are stored in quantum states, whose states are perfectly readable by Alice or any user. Although the classical data is readable by anyone, these classical bits are *encoded* bits, thus Alice or any user needs a *decryption key* to decode the information. The secret quantum information is stored in the rotation angles of the qubits. In the authentication process, the system verifies the decoding quantum key, which is stored on the data medium. The method uses non-orthogonal quantum states, furthermore both secret key states and data states are *non-orthogonal* quantum states (Imre and Balazs, 2005).

### *2.1  Encoding by polarization angle of quantum bits*

In this paper, we will concentrate on the basic idea of the proposed quantum-based data protection system, in the absence of errors caused by technical problems. Although with the implementation of quantum error correcting codes the proposed scheme can be adjusted to fit realistic settings. The quantum states used are non-orthogonal quantum states, the polarization of quantum bits is rotated by a certain angle (Imre and Balazs, 2005).

**Figure 1**  The secret rotation angle of a quantum state

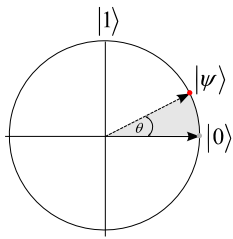

The classical data are stored on the data medium by these non-orthogonal quantum states. If the Issuer stores a classical *n*-bit string, he randomly chooses angle $\theta_i \in [0, 2\pi)$ for every classical bit and prepares the quantum state $|\psi_i\rangle = \cos\theta_i |0\rangle + \sin\theta_i |1\rangle$. The Issuer writes all these *n* quantum states to the data medium, while keeping all $\theta_i \in [0, 2\pi)$ secret. If Alice receives the quantum based data medium from the Issuer, she runs a *simple reading mechanism* to read the classical data on the data medium. For every quantum state, she measures the quantum state $|\psi_i\rangle$ with the projection operator $\mathcal{P} = |0\rangle\langle 0| + |1\rangle\langle 1|$.

The result of the linear operator $\mathcal{P} = |0\rangle\langle 0|$ acting on an unknown $|\psi\rangle = \alpha|0\rangle + \beta|1\rangle$ quantum state, projects the state $|\psi_i\rangle$ into state $|0\rangle$ with probability $|\alpha|^2$, while the linear operator $\mathcal{P} = |1\rangle\langle 1|$ projects into state $|1\rangle$ with probability $|\beta|^2$ (Imre and Balazs, 2005). The components not in the sub-vector space are discarded. Alice can use projector $\mathcal{P}$ to filter out everything other than the subspace she is looking for (Imre and Balazs, 2005). If Alice has a *pirated* data medium, she wants to project a false quantum state

$$|\theta_i^*\rangle = \cos\theta_i^* |0\rangle + \sin\theta_i^* |1\rangle \tag{1}$$

into a valid

$$|\theta_i\rangle = \cos\theta_i |0\rangle + \sin\theta_i |1\rangle \tag{2}$$

state, however she can only do this with a non-zero error rate $\varepsilon$. Alice can make an arbitrary projective measurement in the *XZ* plane, however the projected quantum states can only pass the verification process with probability $(1-\varepsilon)^n$, which becomes *exponentially small* as the length of the string increases. Thus, if Alice has a *pirated* disk, she cannot decode the quantum states correctly, and she gets only useless *garbage* data. The quantum state $|\theta_i\rangle$ can *never be cloned perfectly* because the angle $\theta_i$ is kept secret from Alice and others and it can be varied within the range $[0, 2\pi)$. To conclude the purpose of the proposed system, Alice can retrieve perfectly the encoded *classical data* stored in the quantum states on the data medium without the help of the provider (Brassard et al. 2007).

In Figure 2 we illustrate the situation if Alice sends her angle state to the Issuer, who verifies the angle of the sent quantum state.

**Figure 2** Verification of non-orthogonal quantum state

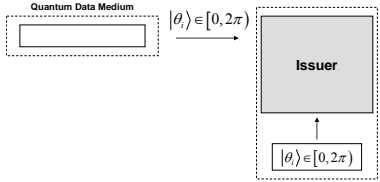

Only the provider can check whether the quantum states on the data medium are original or not, and only the provider knows the decryption key for the *encoded* quantum states. Thus, Alice *gains no benefit* from the knowledge of the classical data stored on the data medium, because these bits are encoded bits and the decoding key is *operating on quantum states*.

## 3   Quantum-based data medium authentication

The *data* states and *secret key* states are encrypted by secret rotation angles. In order to read the disguised quantum bits correctly, Alice must rotate the *i*-th data quantum bit by the secret angle $\theta_i$ in the opposite direction to the Issuer. We can assume for example, that our data state $|d\rangle = a|0\rangle + b|1\rangle$, where $a$ and $b$ are complex coefficients, is a single quantum bit encoded by the rotation operator $\mathcal{R}(\theta_0)$. This rotation operator in matrix form is:

$$\mathcal{R}(\theta_0) = \begin{pmatrix} \cos\theta_0 & \sin\theta_0 \\ -\sin\theta_0 & \cos\theta_0 \end{pmatrix}. \qquad (3)$$

The copy-protection of these rotation angles relies on the no-cloning theorem (Wootters and Zurek, 1982). If a superposition state is measured, the result will be one of two orthogonal states and no information regarding the rotation angle is left. In the authentication method, both the data states, and the data decryption keys are non-orthogonal states. We will describe this method in this paper further on.

The security of the proposed scheme relies on the *no-cloning* theorem (Wootters and Zurek, 1982). Contrary to classical information, in a quantum communication system the quantum information cannot be copied perfectly. If Alice sends a number of photons $|\psi_1\rangle, |\psi_2\rangle, ..., |\psi_N\rangle$ through the quantum channel, an eavesdropper is not interested in copying an arbitrary state, only the possible polarization states. To copy the sent quantum state, an eavesdropper has to use a quantum cloner machine, and a known "*blank*" state $|0\rangle$, onto which the eavesdropper would like to copy Alice's quantum state. If Eve wants to copy the *i*- th sent photon $\left|\psi_i\right\rangle$, she has to apply a unitary transformation $U$, which gives the following result:

$$U\left(\left|\psi_i\right\rangle \otimes \left|0\right\rangle\right) = \left|\psi_i\right\rangle \otimes \left|\psi_i\right\rangle, \tag{4}$$

for each polarization state of qubit $\left|\psi_i\right\rangle$. A photon chosen from a given set of polarization states can only be perfectly cloned if the polarization angles in the set are distinct, and are all mutually orthogonal (Imre and Balazs, 2005). The unknown non-orthogonal states cannot be cloned perfectly, the cloning process of the quantum states is possible only if the information being cloned is classical, hence the quantum states are *all orthogonal*. The polarization states are not all orthogonal states, which makes it impossible for the eavesdropper to copy the sender's quantum states (Brassard, 2006).

*3.1  Brief overview of quantum information processing*

In this section, we give a brief overview of quantum mechanics, and we introduce the basic notation which will be used in the text. In quantum information processing, the logical values of classical bits are replaced by state vectors $\left|0\right\rangle$ and $\left|1\right\rangle$, which is the Dirac notation. Contrary to classical bits, a qubit $\left|\psi\right\rangle$ can also be in a linear combination of basis vectors $\left|0\right\rangle$ and $\left|1\right\rangle$. The state of a qubit can be expressed as $\left|\psi\right\rangle = \alpha\left|0\right\rangle + \beta\left|1\right\rangle$, where $\alpha$ and $\beta$ are complex numbers, this is also called the superposition of the basis vectors, with probability amplitudes $\alpha$ and $\beta$ (Imre and Balazs, 2005). A qubit $\left|\psi\right\rangle$ is a vector in a two-dimensional complex space, where the basis vectors $\left|0\right\rangle$ and $\left|1\right\rangle$ form an orthonormal basis. The orthonormal basis $\{\left|0\right\rangle,\left|1\right\rangle\}$ forms the computational basis (Imre and Balazs, 2005). The vectors or states $\left|0\right\rangle$ and $\left|1\right\rangle$ can be expressed in matrix representation by

$$\left|0\right\rangle = \begin{bmatrix} 1 \\ 0 \end{bmatrix} \text{ and } \left|1\right\rangle = \begin{bmatrix} 0 \\ 1 \end{bmatrix}. \tag{5}$$

If $\left|\alpha\right|^2$ and $\left|\beta\right|^2$ are the probabilities, and the number of possible outputs is only two, then for $\left|\psi\right\rangle = \alpha\left|0\right\rangle + \beta\left|1\right\rangle$ we have $\left|\alpha\right|^2 + \left|\beta\right|^2 = 1$, and the norm of $\left|\psi\right\rangle$ is $\left\|\left|\psi\right\rangle\right\| = \sqrt{\left|\alpha\right|^2 + \left|\beta\right|^2} = 1$. The most general transformation that keeps constraint $\left\|\left|\psi\right\rangle\right\|$ is a linear transformation $U$, that transforms unit vectors into unit vectors. A *unitary* transformation (Nielsen, Chuang, 2000) can be defined as

$$U^\dagger U = UU^\dagger = I, \tag{6}$$

where $U^\dagger = \left(U^*\right)^T$, hence the adjoint is equal to the transpose of the complex conjugate and $I$ is the identity matrix (Imre and Balazs, 2005). The state $|\psi\rangle$ of an *n*-qubit quantum register is a superposition of the $2^n$ states $|0\rangle, |1\rangle, ..., |2^n - 1\rangle$, thus $|\psi\rangle = \sum_{i=0}^{2^n-1} \alpha_i |i\rangle$, with amplitudes $\alpha_i$ constrained to $\sum_{i=0}^{2^n-1} |\alpha_i|^2 = 1$. The state of an *n*-qubit length quantum register is a vector in a $2^n$-dimensional complex vector space, hence if the number of qubits in the quantum register increases linearly, the dimension of the vector space increases exponentially (Nielsen, Chuang, 2000).

A complex vector space $V$ is a Hilbert space $\mathcal{H}$, if there is an *inner product* $\langle \psi | \varphi \rangle$ with $x, y \in \mathbb{C}$ and $|\varphi\rangle, |\psi\rangle, |u\rangle, |v\rangle \in V$ defined by rules of $\langle \psi | \varphi \rangle = \langle \varphi | \psi \rangle^*$, $\langle \varphi | (a|v\rangle + b|v\rangle) \rangle = a \langle \varphi | u \rangle + b \langle \varphi | v \rangle$, and $\langle \varphi | \varphi \rangle > 0$ if $|\varphi\rangle \neq 0$. If we have vectors $|\varphi\rangle = a|0\rangle + b|1\rangle$ and $|\psi\rangle = c|0\rangle + d|1\rangle$, then the inner product in matrix representation can be expressed as

$$\langle \varphi | \psi \rangle = \begin{bmatrix} a^* & b^* \end{bmatrix} \begin{bmatrix} c \\ d \end{bmatrix} = a^* c + b^* d. \tag{7}$$

The outer product between two vectors $|\varphi\rangle$ and $|\psi\rangle$ can be defined as $|\psi\rangle\langle\varphi|$, satisfying $(|\psi\rangle\langle\varphi|)|v\rangle = |\psi\rangle\langle\varphi|v\rangle$ (Imre and Balazs, 2005). The matrix of the outer product $|\psi\rangle\langle\varphi|$ is obtained by the usual matrix multiplication of a column matrix by a row matrix, however the matrix multiplication can be replaced by a tensor product, since $|\varphi\rangle\langle\psi| = |\varphi\rangle \otimes \langle\psi|$. If we have vectors $|\varphi\rangle = a|0\rangle + b|1\rangle$ and $|\psi\rangle = c|0\rangle + d|1\rangle$, the outer product in matrix representation can be expressed as

$$|\varphi\rangle\langle\psi| = \begin{bmatrix} a \\ b \end{bmatrix} \begin{bmatrix} c^* & d^* \end{bmatrix} = \begin{bmatrix} ac^* & ad^* \\ bc^* & bd^* \end{bmatrix}. \tag{8}$$

In Figure 3, we illustrate the general model of an *n*-length quantum register, where the input state $|\psi_i\rangle$ is either $|0\rangle$ or $|1\rangle$, generally. After the application of a unitary transformation $U$ on the input states, the state of the quantum register can be given by state vector $|\psi\rangle$. The unitary operator $U$ is a $2^n \times 2^n$ matrix, with – in principle – an infinite number of possible operators.

**Figure 3**   General sketch of an *n*-length quantum register

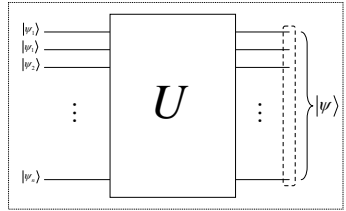

The result of the measurement of state $|\psi\rangle$ results in zeros and ones that form the final result of the quantum computation, based on the *n*-length qubit string stored in the quantum register (Nielsen, Chuang, 2000).

*3.2   Storing information in rotation angles*

We distinguish *data* $|d\rangle = a|0\rangle + b|1\rangle$ qubit states from *secret key*

$$|\psi_\mathcal{K}\rangle = |\theta_1\rangle \otimes |\theta_2\rangle \otimes ... \otimes |\theta_n\rangle \tag{9}$$

qubit states on the data medium. The decryption mechanism is realized by *secret unitary rotations*, therefore the required secret is stored in the rotation angles of the key qubits. Alice takes the data quantum states from the data medium and her *i*-th qubit is valid iff the qubit can be projected into (Acín, Jané and Vidal, 2000), (Kim et al., 2002).

$$\mathcal{R}_\theta |d\rangle = \cos\theta |0\rangle + \sin\theta |1\rangle. \tag{10}$$

Thereby the data string is valid iff all the qubits from the string can be projected successfully. Thus, if Alice wants to copy an original string from the data medium, she has to know all the $\theta_i$, $i = 1,...,n$, secret rotation angles. In the proposed system, Alice has a chance *not greater* than $\varepsilon = \sin^2(\theta_i)$ to read the states exactly from the data medium, because she doesn't know the original rotation angle of the *i*-th qubit $\theta_i$. We would consider approximate those transformations where the output state of the transformation is only an approximation to $\mathcal{R}_\theta |d\rangle = \cos\theta|0\rangle + \sin\theta|1\rangle$. If the angle transformation does succeed, the output state becomes $\mathcal{R}_\theta |d\rangle$. The $p_\theta^d$ *a priori* probability of decryption success depends both on the data quantum state $|d\rangle$ and on the quality of the stored rotation operation $\mathcal{R}_\theta$ (Acín, Jané and Vidal, 2000). Thus, the quality of the proposed quantum-based decryption transformation is quantified by the average of *a priori* probability $\langle p \rangle = \int_{C^2} d(d) \int \frac{d\theta}{2\pi} p_\theta^d$. Since we have only one qubit to encode the rotation operation $\mathcal{R}_\theta$, the angle state $|\theta\rangle = \frac{1}{\sqrt{2}}\left(e^{i\frac{\theta}{2}}|0\rangle + e^{-i\frac{\theta}{2}}|1\rangle\right)$ can be used to store this $\mathcal{R}_\theta$. The decryption method's *control qubit* corresponds to the data qubit $|d\rangle$, while the *target qubit* is equal to the decryption key state $|\theta\rangle$. Thus, using a simple CNOT transformation (Nielsen, Chuang, 2000), the state is transformed to $|d\rangle \otimes |\theta\rangle \to \frac{1}{\sqrt{2}}\left(\mathcal{R}_\theta |d\rangle \otimes |0\rangle + \mathcal{R}_\theta^\dagger |d\rangle \otimes |1\rangle\right)$, and therefore a projective measurement in the $\{|0\rangle, |1\rangle\}$ basis of the program register will make the data qubit collapse either into the desired state $\mathcal{R}_\theta |d\rangle$ or into the wrong state $\mathcal{R}_\theta^\dagger |d\rangle$ (Acín, Jané and Vidal, 2000).

## 4 Data states and key states on data medium

We store unitary operation $\mathcal{R}_{\theta_i}$ in the rotation angle $|\theta_i\rangle$. This transformation can be performed on a data qubit by an arbitrary system almost perfectly. In the proposed system, we use *data* states $|d_1\rangle \otimes \ldots \otimes |d_n\rangle$, and decryption *key* states $|\psi_{\mathcal{K}_{D,1}}\rangle \otimes \ldots \otimes |\psi_{\mathcal{K}_{D,n}}\rangle$, where $|\theta_i\rangle = \mathcal{R}_{\theta_i}$ and the $i$-th bit of the key operates on $|d_i\rangle = a_i |0\rangle + b_i |1\rangle$, where $a_i$ and $b_i$ are complex coefficients (Acín, Jané and Vidal, 2000). The $i$-th key transformation is applied on $|d_i\rangle$, so the manipulation of the joint state $|d_i\rangle \otimes |\theta_i\rangle$ does not require knowing the operation $\mathcal{R}_{\theta_i}$ nor the data state $|d_i\rangle$. Thus, Alice or an arbitrary user who has a valid quantum protected data medium is able to transform the data and key state into $\mathcal{R}_{\theta_i} |d_i\rangle \otimes |\chi_{d,\mathcal{R}(\theta_i)}\rangle$, where $|\chi_{d,\mathcal{R}(\theta_i)}\rangle$ is just some residual state. Thus, if Alice reads the data medium, the result of the action $\mathcal{R}_{\theta_i}$ on an arbitrary data state $|d_i\rangle$ is determined by the secret decryption key state $|\theta_i\rangle$ (Acín, Jané and Vidal, 2000).

Let the secret, one-qubit key state unitary operation be $\mathcal{R}_\theta \equiv e^{(i\theta\sigma_z/2)}$, for an arbitrary angle $\theta \in [0, 2\pi)$, which corresponds to an arbitrary rotation around the $\hat{z}$-axis of a spin ½ particle (Nielsen, Chuang, 2000). The *secret key state* can also be given by $|\theta\rangle = \frac{1}{\sqrt{2}}\left(e^{i\theta}|0\rangle + e^{-i\theta}|1\rangle\right)$, which is prepared by the Issuer, applying $\mathcal{R}_\theta$ on the qubit in the standard superposition state. The Issuer also prepares and writes to the data medium another qubit $|d\rangle = a|0\rangle + b|1\rangle$, along with the key state $|\theta\rangle$.

**Figure 4** The Issuer puts data state and key quantum state on the quantum disk

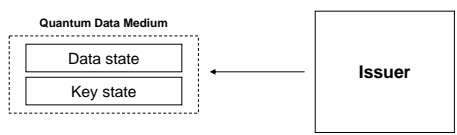

If Alice buys a valid data medium from a store and she tries to read the quantum-protected data medium at home, she doesn't know the secret key $|\theta\rangle$ nor the complex coefficients $a$ and $b$. However she can read the data medium successfully, because she is able to prepare the decrypted state $\mathcal{R}_\theta|d\rangle$ from the unknown projected but still unmeasured encrypted data and the projected but still unmeasured unknown secret key.

The readout process of the superposition quantum states and the encryption process might be controlled by Alice's *quantum-CPU* with a pre-prepared *Quantum Instruction List* stored on the medium. Alice's encryption process is defined only in the angles of quantum states in the Hilbert space $\mathcal{H}$. The secret rotations are implemented by a quantum-CPU in Alice's reader device with a *data* and *program register*. The data register contains the readout data quantum states, while the program register contains the readout key states. The device's quantum processing unit uses the readout non-orthogonal quantum states stored in the data and program register (Nielsen, Chuang, 2000).

**Figure 5** The readout process and the execution of the readout quantum program on the data state

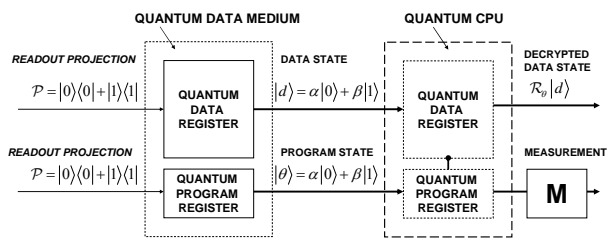

The quantum CPU's data register contains the readout data quantum states, while the program register contains the readout key states. The device's quantum processing unit uses the readout non-orthogonal quantum states stored in the data and program register.

*4.1  Decrypting unknown data with unknown secret key*

In order to implement the *unknown* unitary transformation $\mathcal{R}_\theta$ on *unknown* data state $|d\rangle$, Alice has to perform a CNOT operation (Nielsen, Chuang, 2000), taking the data qubit in a valid state $|d\rangle$ as the *control-qubit* and the secret key qubit in state $|\theta\rangle$ as the *target-qubit*. The CNOT gate $|0\rangle\langle 0| \otimes I + |1\rangle\langle 1| \otimes \sigma_x$, where $\sigma_x$ is the Pauli *X*-transformation (Imre and Balazs, 2005), permutes the $|0\rangle$ and $|1\rangle$ states of the target only if the control qubit is in state $|1\rangle$ (Acín, Jané and Vidal, 2000). The two-qubit state is transformed to

$$|d\rangle \otimes |\theta\rangle \rightarrow \frac{1}{\sqrt{2}}\left(\mathcal{R}_\theta |d\rangle \otimes |0\rangle + \mathcal{R}_\theta^\dagger |d\rangle \otimes |1\rangle\right). \tag{11}$$

Therefore, a projective measurement on the basis $\{|0\rangle, |1\rangle\}$ of the decryption key state will make the data qubit collapse either into the desired state $\mathcal{R}_\theta |d\rangle$ or into the wrong state $\mathcal{R}_\theta^\dagger |d\rangle$, with each outcome having probability of 1/2 (Nielsen, Chuang, 2000). If Alice obtains the wrong result $\mathcal{R}_\theta^\dagger |d\rangle$, then she fails at performing the wished operation and will be unable to read the data medium correctly. In order to construct a more efficient solution, we use multi-qubit secret key states (Acín, Jané and Vidal, 2000), (Kim et al., 2002).

Let us assume that Alice measures the *i*-th qubit in the basis $\{|0\rangle, |1\rangle\}$ and gets the data state $|d_i^*\rangle = a^* |0\rangle + b^* |1\rangle$, where $a^*$ and $b^*$ are complex coefficients. Due to the no-cloning theorem of quantum states (Wootters and Zurek, 1982), Alice cannot determine and copy the state exactly since she does not know the angle $\theta_i$. Thus, she cannot be sure whether the *i*-th readout quantum state $|d_i^*\rangle$ is identical to the original data state $|d_i\rangle$ or not. Thus, if Alice tries to copy the state, she picks up another angle $\theta_i$ and she prepares a fake state $|\theta_i^*\rangle = \cos\theta_i^* |0\rangle + \sin\theta_i^* |1\rangle$. Now, Alice has a chance that $|d_i^*\rangle = |d_i\rangle$ with probability $\cos^2(\theta_i)$, and has a chance that $|d_i^*\rangle \neq |d_i\rangle$ with probability $\sin^2(\theta_i)$. Therefore, a fake state $|\theta_i^*\rangle$ can be projected to the original state $|\theta_i\rangle = \cos\theta_i |0\rangle + \sin\theta_i |1\rangle$ successfully with probability

$$p_i = \cos^2\theta_i \cos^2\left(\theta_i - \theta_i^*\right) + \sin^2\theta_i \sin^2\left(\theta_i + \theta_i^*\right). \tag{12}$$

In the proposed system, the rotation angles are evenly distributed and the *maximum of the average probability* $\bar{p}_i$ can be reached if the angle of state $\theta_i^*$ is equal to zero (Acín, Jané and Vidal, 2000). Thus, the best she can do after she has measured the qubit which is stored on the data medium is to leave it as it is after the measurement. In this case, she gets the correct $\theta_i$ with probability

$$p_i = 1 - \frac{1}{2}\sin^2 2\theta_i, \tag{13}$$

therefore Alice's total probability of coping with $n$ quantum states with valid rotation angles $\theta_i$ is

$$P = \prod_{i=1}^{n}\left(1 - \frac{1}{2}\sin^2 2\theta_i\right). \tag{14}$$

Thus, by increasing *n*, Alice's probability of success can be made arbitrarily small (Acín, Jané and Vidal, 2000).

*4.2 One-qubit secret key state realized by multi-qubit string*

In this section we show the data storage and decryption method of the proposed system. The one-qubit secret key state $|\theta\rangle = \cos\theta|0\rangle + \sin\theta|1\rangle$ is realized by a multi-qubit string, which is stored on the data medium. If Alice fails to decode the data qubit, a second go can correct the wrong state $\mathcal{R}_\theta^\dagger|d\rangle$ into the right state $\mathcal{R}_\theta|d\rangle$ (Nielsen, Chuang, 2000).

Therefore, Alice needs to apply the gate in Figure 6 to prepare the *bad state* $\mathcal{R}_\theta^\dagger|d\rangle$ or the right state $\mathcal{R}_\theta|d\rangle$ with equal probability 1/2. If Alice wants to increase her probability of success, she has to insert a new key angle state $|2\theta\rangle$, which is prepared and stored on the data medium by the Issuer. To prepare the state $|2\theta\rangle$, the Issuer needs only to perform the operation $\mathcal{R}_\theta$ twice on a qubit in a standard superposition state (Acín, Jané and Vidal, 2000), (Kim et al., 2002). The Issuer writes the key state $|\theta\rangle \otimes |2\theta\rangle$ to the data medium in order to get the right decrypted state $\mathcal{R}_\theta|d\rangle$. Thus, if the $|2\theta\rangle$ state is available on the data medium for Alice's second attempt, she can perform the right $\mathcal{R}_\theta$ transformation on data qubit $|d\rangle$ with probability 3/4.

**Figure 6** The decryption of a single-qubit data with single-qubit key state. The secret angle state is stored in one qubit

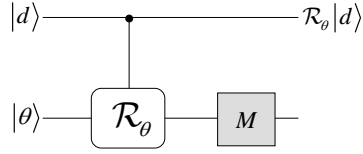

In case of a new failure, the state of the system becomes $\mathcal{R}_\theta^{\dagger 3}|d\rangle$. The improved gate in Figure 7 contains a *Toffoli*-gate, which acts as a CNOT-gate (Nielsen, Chuang, 2000) between the first and third line of the circuit only if the second line carries a $|1\rangle$ state. If the second line carries $|1\rangle$, this means that the previous measurement on key state $|\theta\rangle$ failed. In the case of a new failure, Alice can correct it by inserting this state again, together with state $|4\theta\rangle$ (Acín, Jané and Vidal, 2000).

**Figure 7** The decryption of a single-qubit data with multi-qubit key states. The secret angle state is stored in two quantum states

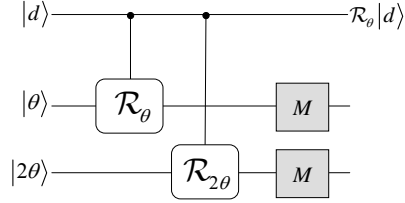

If Alice has no luck and keeps on failing, she can try to correct the state as many times as she wishes, provided that the state $|2^{l-1}\theta\rangle$ is available on the data medium for the $l$-th attempt. Therefore, an $l$-qubit state $\otimes_{i=1}^{l}|2^{i-1}\theta\rangle$ can be used to implement the secret transformation $\mathcal{R}_\theta$ on the corresponding data qubit $|d\rangle$ with probability $1-(1/2)^l$. Thus, Alice takes data qubit $|d\rangle$ as the control bit, and takes decrypting qubit states $\otimes_{i=1}^{l}|2^{i-1}\theta\rangle$ as the target, therefore Alice's reader device evolves the transformation of $|d\rangle \otimes_{i=1}^{l}|2^{i-1}\theta\rangle$ into

$$\frac{1}{\sqrt{2^l}}\left(\sqrt{2^l-1}\,\mathcal{R}_\theta\,|d\rangle \otimes |right\rangle + \mathcal{R}_\theta^{(2^l-1)\dagger}|d\rangle \otimes |wrong\rangle\right), \qquad (15)$$

and a *posterior* measurement of the secret key states either in the $|right\rangle$ or $|wrong\rangle$ state, where $\langle right|wrong\rangle = 0$. Since the Issuer gives an *l*-length key $\otimes_{i=1}^{l}|2^{i-1}\theta\rangle$ to decode data state $|d\rangle$, Alice's probability of failure will be $\varepsilon = (1/2)^{l}$ and it decreases exponentially with the size of the $|\theta\rangle$ decrypting state's secret string length $l$ (Acín, Jané and Vidal, 2000), (Kim et al., 2002). This gate is shown in Figure 8.

**Figure 8** The decryption of single-qubit data with multi-qubit key state. The secret angle state is stored in an *l*-length quantum string

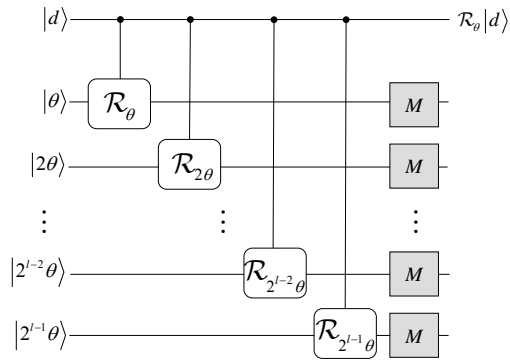

In Figure 9, Alice's simplified quantum circuit is equivalent to the previous gate, since it takes the unknown data state $|d\rangle$ as the control, and the unknown secret key state $|\theta\rangle$ as the target of the CNOT-gate (Nielsen, Chuang, 2000). Depending on the result of a measurement on the key qubit state, Alice produces *either* $\mathcal{R}_\theta|d\rangle$ *or* $\mathcal{R}_\theta^\dagger|d\rangle$.

**Figure 9** The decryption of single-qubit data with key qubit state. The short diagonal line on the bottom line indicates that the secret key bit state consists of several quantum bits

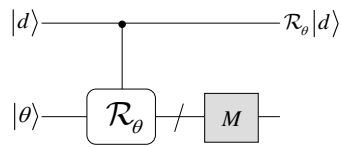

The verification uses an *l*-qubit for the realization of every secret $|\theta\rangle$ rotation angle. Thus, every rotation transformation $\mathcal{R}_\theta$ succeeds with probability $p = 1 - (1/2)^l$, with error probability $\varepsilon = (1/2)^l$ (Nielsen, Chuang, 2000). Since the Issuer writes only a one-state decryption key to the data medium, Alice fails to perform $\mathcal{R}_\theta |d\rangle$ with probability $p_1 = 1/2$. If the Issuer writes a two-qubit decryption key state to the data medium, Alice fails with probability $p_2 = 1/4$, etc. The average length of the required string is

$$\bar{l} = \sum_{l=1}^{\infty} p_l l = \sum_{l=1}^{\infty} \frac{l}{2^l} = 2. \tag{16}$$

Thus, a two-qubit state for a single-qubit data is sufficient, on average (Acín, Jané and Vidal, 2000), (Kim et al., 2002).

### *4.4 Entropy of secret angle states*

We have seen that, on average, a two-qubit state is sufficient to store and retrieve any angle state. In this Section we step forward, and we claim that we can store the unitary operation $\mathcal{R}_\theta$ with arbitrary, unknown $\theta$ angle using on average less than four quantum bits, if $0 \leq \theta \leq \pi$. The rotation operator $\mathcal{R}_\theta$ for arbitrary $\theta$ can be implemented by a sequence of operations of the form $\mathcal{R}_{\theta_k}$ with binary angles $\theta_k = \pi/2^k$. Thus, we can consider the implementation of $\mathcal{R}_{\theta_k}$ for a certain $k = N$ and $\theta_N = \pi/2^N$. The required $\rho_N$ set of $N$ states is $\rho_N = \{|\mathcal{R}_{\theta_N}\rangle, |\mathcal{R}_{2\theta_N}\rangle, \ldots |\mathcal{R}_{2^{N-1}\theta_N}\rangle\}$, where the corresponding probabilities are given by $p_l = 1/2^{l-1}$ for the $l^{th}$ state. If Alice retrieves the original angle state $\theta$, each step can be considered independently and involves with probability $p = 1/2$ either the storage of the state $|\mathcal{R}_{2^{N-1}\theta_N}\rangle$ for the $l^{th}$ step (Acín, Jané and Vidal, 2000).

The entropy is $\mathbf{S}_j = -x_j \log_2(x_j) - (1 - x_j)\log_2(1 - x_j)$, where $x_j = (1 + \cos\theta_j)/2$, and $\theta_j = \pi/2^j$. That is, the total number of quantum states required to store and retrieve the angle $|\theta\rangle$ or the operation $|\mathcal{R}_{\theta_N}\rangle$ is given by $\sum_{l=1}^{N} \mathbf{S}_{N-l} \frac{1}{2^{l-1}}$. The total number of quantum states needed to store and retrieve the angle state, with arbitrary $\theta$, $0 \leq \theta \leq \pi$, for $1 \leq k \leq \infty$, is on average given by $\sum_{k=1}^{\infty} \mathbf{S}_k \sum_{l=0}^{k-1} \frac{1}{2^l} \leq 2 \sum_{k=1}^{\infty} \mathbf{S}_k \leq 4$. That is, less than four quantum states per angle state are required on average to store an arbitrary, unknown operation of the form $\mathcal{R}_\theta$, and in the previous sections we have seen that on average two qubits suffice to implement $\mathcal{R}_\theta$ (Acín, Jané and Vidal, 2000).

*4.5 Distribution of quantum copy protected data medium*

Alice will be able to obtain the result of the *l-length* decryption operation $\mathcal{R}_\theta^{(l)}$ and the *n-length* decrypted data state $\mathcal{R}_\theta^{(l)}\left|d^{(n)}\right\rangle$ only if the Issuer puts at least a two state length string $\otimes_{i=1}^2 \left|2^{i-1}\theta\right\rangle = \left|\theta\right\rangle \otimes \left|2\theta\right\rangle$ on the data medium for decryption.

This means that Alice, or anyone who buys an original quantum copy-protected data medium, has to be able to compute the secret decrypting unitary operator $\mathcal{R}_\theta^{(l)}$ on an *n*-qubit data string $\left|d^{(n)}\right\rangle$. The Issuer puts both data qubits $\left|d^{(n)}\right\rangle$ and key states $\left|\theta^{(l)}\right\rangle$ on the data medium, and Alice operates the decrypting rotation $\mathcal{R}_\theta^{(l)}\left|d^{(n)}\right\rangle$, but she has no information about what the secret unitary operation $\mathcal{R}_\theta^{(l)}$ was (Acín, Jané and Vidal, 2000). Thus, the Issuer doesn't want Alice to know what secret decrypting algorithm she is running on her reader device. The secret key states of the *l-length* state $\left|\theta^{(l)}\right\rangle$ are *non-orthogonal* states, because Alice can in principle distinguish between orthogonal states (Nielsen, Chuang, 2000). If Alice tries to copy $\left|\theta^{(l)}\right\rangle$, then she will necessarily modify the key state, which will result in an incorrect decrypted state (Acín, Jané and Vidal, 2000). That is, it is not possible for Alice to copy, even in an approximate form, the secret key state or to decrypt the data states. An *N*-qubit length unknown state $\left|\mathcal{R}_\theta^N\right\rangle$ has maximal entropy, thus $\int \frac{d\theta}{2\pi}\left|\mathcal{R}_\theta^N\right\rangle\left\langle\mathcal{R}_\theta^N\right| = \left(\frac{\mathbf{1}}{2}\right)^{\otimes N}$, where $\left(\frac{\mathbf{1}}{2}\right)$ is the maximally mixed state (Nielsen, Chuang, 2000). Let assume $p_\theta$ be the probability of a successful transformation $\mathcal{R}_\theta\left|d\right\rangle = \left|\mathcal{R}_\theta\right\rangle$, which is independent of data state $\left|d\right\rangle$. The quantity to be maximized is $\left\langle p\right\rangle = \int \frac{d\theta}{2\pi}p_\theta$, the *average probability* of success (Acín, Jané and Vidal, 2000). If we choose a random angle $\theta_0 \in [0,2\pi)$ and transform the state $\left|\mathcal{R}_\theta^N\right\rangle$, our state will change to $\left|\mathcal{R}_{\theta+\theta_0}^N\right\rangle$. This can be achieved if the state $\left|\mathcal{R}_\theta^N\right\rangle$ is transformed by $\mathcal{R}_{\theta_0}^1 \otimes \mathcal{R}_{\theta_0}^2 \otimes \mathcal{R}_{\theta_0}^3 \otimes ... \otimes \mathcal{R}_{\theta_0}^{N-1}$, we then obtain $\mathcal{R}_{\theta+\theta_0}|d\rangle$. We have the state $\mathcal{R}_{\theta+\theta_0}|d\rangle$ and finally we eliminate $\mathcal{R}_{\theta_0}$ by performing $\mathcal{R}_{\theta_0}^\dagger$ on $\mathcal{R}_{\theta+\theta_0}|d\rangle$ (Acín, Jané and Vidal, 2000). Let us assume that our data state $|d\rangle = |\beta\rangle$, so that $\mathcal{R}_\theta|\beta\rangle = |\theta+\beta\rangle$. The transformation is unitary, thus if $\beta^* = \pi + \beta + \theta - \theta^*$, then $\langle\beta^*|\beta\rangle\langle\mathcal{R}_{\theta^*}^N|\mathcal{R}_\theta^N\rangle = (1-p)\langle\chi_{\theta^*}^\beta|\chi_\theta^\beta\rangle$, and $\langle\beta^*|\mathcal{R}_{\theta^*}^\dagger \mathcal{R}_\theta \beta\rangle = 0$ (Acín, Jané and Vidal, 2000).

In Figure 10 we show the probability success as a function of the length of the key qubit string. The vertical axis represents the probability of success, the horizontal axis shows the length of the input qubit string.

**Figure 10** Probability of success of decryption process as a function of length of key qubit string

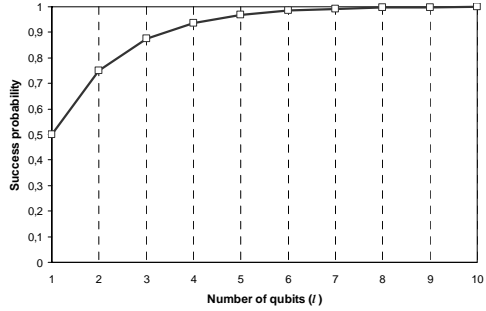

If we take $\theta - \theta^* = \pi(1/2)^N$ then $p \leq 1 - (1/2)^N$, and this is what we wanted to show. Thus, if retrieving an angle state $|\theta\rangle$ encoded in $N$ qubits as $|\mathcal{R}_\theta^N\rangle$ the method necessarily fails with probability $\varepsilon = (1/2)^N$ (Acín, Jané and Vidal, 2000).

*4.6   The hash quantum state*

Let us assume that the secret decryption key $\mathcal{K}_D$ and the $i$-th data state are stored on the quantum-based data medium in quantum states $|\psi_{\mathcal{K}_D}\rangle$ and $|d_i\rangle$. Alice has to generate a single-qubit *hash* state from the states of the decryption key and then she has to apply it to a fixed data qubit. Finally, she has to compare it with the validated result.

The secret key $|\psi_{\mathcal{K}_D}\rangle = |\theta_1\rangle \otimes |\theta_2\rangle \otimes ... \otimes |\theta_n\rangle$ is an $n$-bit length string, where the $i$-th bit of the key is $|\theta_i\rangle = \cos\theta_i|0\rangle + \sin\theta_i|1\rangle$. Only the Issuer knows the secret

$\theta_i \in [0, 2\pi)$ rotation angle, thus only the Issuer is able to produce the right hash state $f_{\mathcal{H}}(|\psi_{\mathcal{K}_D}\rangle)$ from these rotation angles. The hash quantum state $f_{\mathcal{H}}(|\psi_{\mathcal{K}_D}\rangle)$ of the secret key is a single qubit state, and it can be generated using $f_{\mathcal{H}}(|\psi_{\mathcal{K}_D}\rangle) = \mathcal{R}_{\theta_1} \cdot \mathcal{R}_{\theta_2} \cdot \ldots \cdot \mathcal{R}_{\theta_n} |d\rangle$. Here $\mathcal{R}_{\theta_i}$ is the rotation operator, which is realized by the *i*-th quantum state $|\theta_i\rangle$ of the decryption key, and $|d\rangle$ is the fixed data state for the verification process. For simplicity, here we assume both the data quantum state and resulting hash quantum state are single qubit states (Nielsen, Chuang, 2000). In Figure 11, we show the case if the original hash state is stored by the Issuer. After the Issuer receives the computed hash state $\mathcal{R}_{\theta_1} \cdot \mathcal{R}_{\theta_2} \cdot \ldots \cdot \mathcal{R}_{\theta_n} |d\rangle$, he compares it with the original, stored hash state $f_{\mathcal{H}}(|\psi_{\mathcal{K}_D}\rangle)$.

**Figure 11**  Generation and verification of hash quantum state

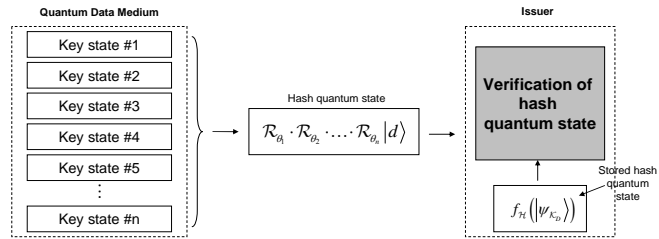

The secret decryption key $|\psi_{\mathcal{K}_D}\rangle$ contains $n$ quantum states $|\theta_1\rangle \otimes |\theta_2\rangle \otimes \ldots \otimes |\theta_n\rangle$, where every $\mathcal{R}_{\theta_i}$ rotation is realized by an *l*-length multi-qubit string, therefore every secret transformation $\mathcal{R}_{\theta_i}$ on the corresponding data qubit $|d\rangle$ can be implemented with probability of success $1 - (1/2)^l$. The elements of the quantum *medium* authentication process – the secret key $|\psi_{\mathcal{K}_D}\rangle$, the valid hash of the key $f_{\mathcal{H}}(|\psi_{\mathcal{K}_D}\rangle)$ and the data state for the verification process $|d\rangle$ – are stored on the data medium. In the verification process, Alice first chooses the corresponding data state $|d\rangle$. She uses this state as the control of a CNOT-gate (Nielsen, Chuang, 2000). The target consists of all of the states of the key $|\psi_{\mathcal{K}_D}\rangle$. In order to implement a successful verification process, Alice has to generate a valid hash state $f_{\mathcal{H}}(|\psi_{\mathcal{K}_D}\rangle)$ from the secret key states. First, Alice takes the data state $|d\rangle$ for the verification and applies the first transformation $\mathcal{R}_{\theta_1}$.

After she measures $|\theta_1\rangle$, she takes the next secret key qubit $|\theta_2\rangle$, and applies $\mathcal{R}_{\theta_2}$ on the state $\mathcal{R}_{\theta_1}|d\rangle$, and so on.

**Figure 12** Generation process of hash quantum angle state. Every angle state rotates the previous state by a certain angle. After the rotations, the Issuer identifies the quantum data medium

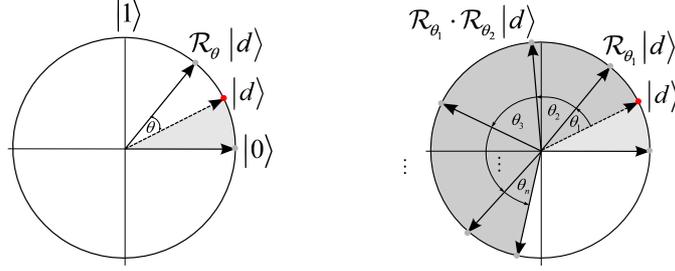

We assume that the length of the key is *n*, so she finally takes the *n*-th qubit $|\theta_n\rangle$ of the key, and she gets the appropriate state $\mathcal{R}_{\theta_1} \cdot \mathcal{R}_{\theta_2} \cdot \ldots \cdot \mathcal{R}_{\theta_n}|d\rangle$ which is equal to $f_{\mathcal{H}}\left(|\psi_{\mathcal{K}_D}\rangle\right)$ iff all of the key states are valid. In Figure 13, we show the preparation of the hash state using a single-qubit data qubit $|d\rangle$.

**Figure 13** The preparation of the hash state on single-qubit data. The *i*-th rotation operator is realized by the key state which is stored in a multi-qubit string

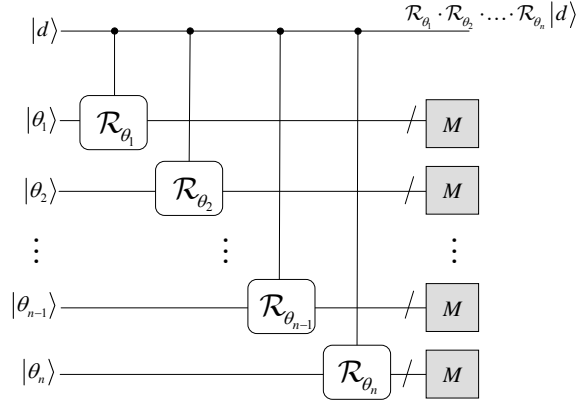

Once the hash state $\mathcal{R}_{\theta_1} \cdot \mathcal{R}_{\theta_2} \cdot \ldots \cdot \mathcal{R}_{\theta_n}|d\rangle$ is generated from the secret key and the fixed data qubit, Alice uses a simple *controlled-SWAP* gate (Nielsen, Chuang, 2000). The two inputs of the controlled-SWAP gate are the generated hash state

$\mathcal{R}_{\theta_1} \cdot \mathcal{R}_{\theta_2} \cdot \ldots \cdot \mathcal{R}_{\theta_n} |d\rangle$ and the original hash state $f_{\mathcal{H}}(|\psi_{\mathcal{K}_D}\rangle)$, which is stored on the data medium. The gate is shown in Figure 14.

**Figure 14** The controlled-SWAP gate (Nielsen, Chuang, 2000) test to check the prepared hash state

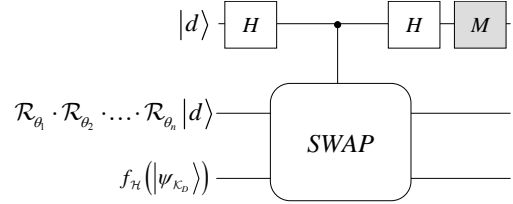

If we assume that the two inputs $|\phi\rangle$ and $|\psi\rangle$ are equal, then the output of the gate is 0, with probability $\Pr[0] = \frac{1}{2}(\langle\phi,\psi| + \langle\psi,\phi|)(|\phi,\psi\rangle + |\psi,\phi\rangle) = \frac{1}{2} + \frac{1}{2}|\langle\psi|\phi\rangle|^2$. If it was the case that $|\phi\rangle = |\psi\rangle$, then we know that this answer is always correct because we will always measure 0. In Figure 15, the original hash state is denoted by $|\psi\rangle = f_{\mathcal{H}}(|\psi_{\mathcal{K}_D}\rangle)$, the hash for verification process – computed from key angle states – is denoted by $|\phi\rangle = \mathcal{R}_{\theta_1} \cdot \mathcal{R}_{\theta_2} \cdot \ldots \cdot \mathcal{R}_{\theta_n} |d\rangle$. The angle between hash states $|\psi\rangle$ and $|\phi\rangle$ determines the probability of success of the SWAP test (Nielsen, Chuang, 2000).

**Figure 15** Distance between valid and invalid quantum state from a pirated disk

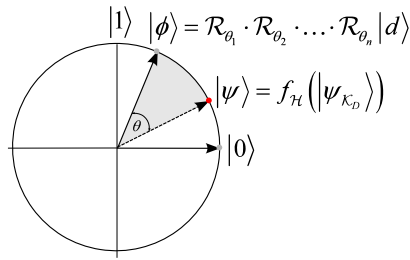

In Figure 16, we illustrate the probability of success of a SWAP test-based (Nielsen, Chuang, 2000) hash state verification process. The vertical axis represents the probability of success of verification, the horizontal-axis represents the difference between the generated and original hash states. As can be concluded, if the hash states are equal $|\phi\rangle = |\psi\rangle$, then the SWAP test will determine the validity with probability of success 1. Otherwise, if the two states are orthogonal to each other, then the result of the test becomes fully quantum probabilistic (Imre and Balazs, 2005), and if the Issuer measures 1, the received hash state will be rejected. In Figure 16, we show the probability of success for a one-qubit length quantum system. The probability of success of the SWAP test (Nielsen, Chuang, 2000) increasing exponentially with the number of qubits in the hash string increases linearly (Acín, Jané and Vidal, 2000).

**Figure 16** Probability of success of SWAP test as a function of the distance between the valid hash state and invalid quantum state for a one-qubit length environment

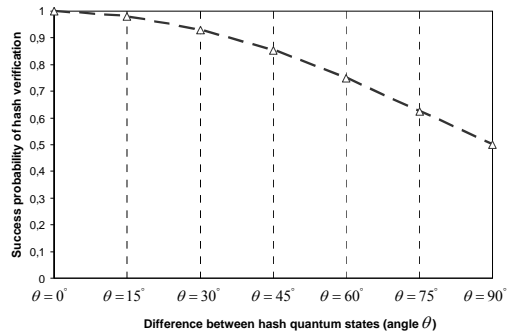

As can be concluded, for an *n*-length quantum system, the probability of success of the test becomes $\Pr[0] = 1 - (1/2)^n$.

### 4.7 Maximum distance between indistinguishable angle states

We can derive the degree of orthogonality between the $|\mathcal{R}_\theta\rangle$ states, with error parameter $\varepsilon$. Using the general decryption operator $\mathcal{G}$, we get $\mathcal{G}[|d\rangle \otimes \mathcal{R}_\theta \otimes |0\rangle] = \sqrt{p_\theta^d}\left(\mathcal{R}_\theta |d\rangle\right) \otimes |\tau_\theta^d\rangle + \sqrt{1-p_\theta^d}|\chi_\theta^d\rangle$, where $p_\theta^d = (1-\varepsilon)$ and $1 - p_\theta^d = \varepsilon$. The state $\mathcal{R}_\theta |d\rangle$ only depends on $|d\rangle$ through a term of order $\varepsilon^\tau$, where $\tau = 1$ for probabilistic gates (Nielsen, Chuang, 2000). Furthermore, for any two data states $|d_1\rangle$ and $|d_2\rangle$ the *inner product* is given by $\langle d_1 | d_2 \rangle = \langle d_1 | d_2 \rangle \langle \mathcal{R}_\theta^{d_1} | \mathcal{R}_\theta^{d_2} \rangle + \mathcal{O}(\varepsilon^\tau)$, where the term $\mathcal{O}(\varepsilon^\tau)$ is linear in $\varepsilon^\tau$. Thus, $\langle \mathcal{R}_\theta^{d_1} | \mathcal{R}_\theta^{d_2} \rangle = 1 + \mathcal{O}(\varepsilon^\tau)$, that is $|\mathcal{R}_\theta^d\rangle = |\mathcal{R}_\theta\rangle + \mathcal{O}(\varepsilon^\tau)$. Let us consider that we have an arbitrary $|d\rangle$ quantum data state, and two unitary

angles $\mathcal{R}_1$ and $\mathcal{R}_2$, thus $\langle \mathcal{R}_1 | \mathcal{R}_2 \rangle = \langle d | R_1^\dagger R_2 | d \rangle \langle \mathcal{R}_1 | \mathcal{R}_2 \rangle + \mathcal{O}(\varepsilon^\tau)$. The scalar product $\langle \mathcal{R}_1 | \mathcal{R}_2 \rangle$ does not depend on the quantum data $|d\rangle$, therefore the dependence of $\langle d | R_1^\dagger R_2 | d \rangle \langle \mathcal{R}_1 | \mathcal{R}_2 \rangle$ on $|d\rangle$ must be of order $\mathcal{O}(\varepsilon^\tau)$ at most (Acín, Jané and Vidal, 2000). Let us assume that $\mathcal{D}_{\mathcal{R}_1, \mathcal{R}_2}$, the distance between angles $\mathcal{R}_1$ and $\mathcal{R}_2$, is $\mathcal{D}_{\mathcal{R}_1, \mathcal{R}_2} = \sqrt{\left(Tr(\mathcal{R}_1 - \mathcal{R}_2)^\dagger (\mathcal{R}_1 - \mathcal{R}_2)\right)}$. If the two angles $\mathcal{R}_1$ and $\mathcal{R}_2$ are very close, the largest variation of $\langle d | R_1^\dagger R_2 | d \rangle$ in $\langle d | R_1^\dagger R_2 | d \rangle \langle \mathcal{R}_1 | \mathcal{R}_2 \rangle + \mathcal{O}(\varepsilon^\tau)$ for two different qubits is $\langle 1 | R_1^\dagger R_2 | 1 \rangle - \langle n | R_1^\dagger R_2 | n \rangle = \theta_1 - \theta_n = \theta_1 + |\theta_n|$, thus $\theta_1 + |\theta_n| \geq \sqrt{\sum_i |\theta_i^2|} / n = \mathcal{D}_{\mathcal{R}_1, \mathcal{R}_2} / \sqrt{2}n$. From this result, the angle states $\mathcal{R}_1$ and $\mathcal{R}_2$ are indistinguishable only if $\mathcal{R}_1$ and $\mathcal{R}_2$ are also very close to each other, thus $\mathcal{D}_{\mathcal{R}_1, \mathcal{R}_2} \ll 1$. If Alice is a dishonest user, she is unable to distinguish between them (Acín, Jané and Vidal, 2000).

*4.8   Security of medium verification method*

In summary, if Alice is a malevolent user and she wants to pass the verification process with a modified pirated disk, then:

- she has to identify all the secret rotation angles $\theta_1$ to $\theta_n$ of the key which is impossible according to the no-cloning theorem (Wootters and Zurek, 1982), or she has to generate a valid $f_\mathcal{H}(|\psi_{\mathcal{K}_D}\rangle)$ hash state from an invalid key $|\psi_{\mathcal{K}_D^*}\rangle$, which is also impossible
- she has to try to use an invalid hash state together with an invalid key. In that case, the verification process accepts the invalid $f_{\mathcal{H}^*}(|\psi_{\mathcal{K}_D^*}\rangle)$ hash state with the fake key. The data states on the disk with fake key $|\psi_{\mathcal{K}_D^*}\rangle$ will be indecipherable.

*4.9   Multi-qubit decryption*

After the verification process has succeeded, Alice can use the states of the decryption key $|\psi_{\mathcal{K}_D}\rangle$ to rotate the encrypted data states in *one-step*, as shown in Figure 17.

**Figure 17**   Decryption of an *n*-length multi-qubit data string with an *n*-length multi-qubit *key-string*. Each state of the key is realized by an *l*-length multi-qubit string. The *i*-th secret angle state operates on the *i*-th data qubit

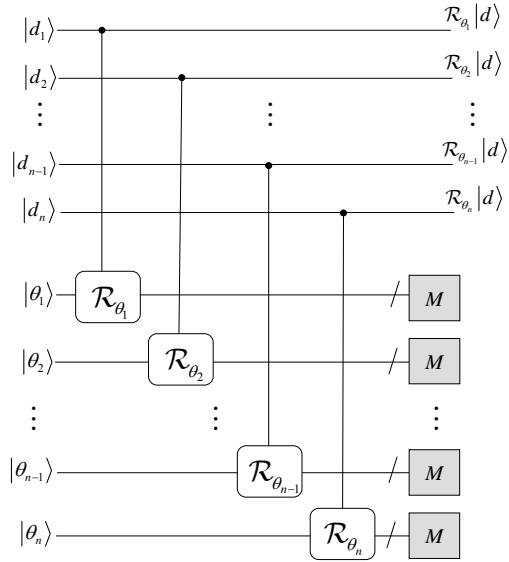

If Alice has an *n*-bit length encrypted data string $|d_1\rangle \otimes ... \otimes |d_n\rangle$, she can decode it with the *verified* *n*-bit length key $|\psi_{\mathcal{K}_D}\rangle = |\theta_1\rangle \otimes |\theta_2\rangle \otimes ... \otimes |\theta_n\rangle$ in one-step. The decryption process can be realized by simple rotation transformations, hence the encrypted data states can be deciphered with high probability and high efficiency. All the rotation transformations are realized by an *l*-length multi-qubit string, therefore every secret transformation $\mathcal{R}_{\theta_i}$ on the corresponding data qubit $|d_i\rangle$ can be implemented with probability of success $1 - (1/2)^l$.

## 5   An experimental realization of quantum protected data medium

In this section, we show one possible experimental realization of the proposed quantum copy protection scheme. Holographic data storage is a potential technology to realize quantum data medium. In holographic technologies, each bit is stored as optical changes on the surface of the holographic data disk. Moreover, holographic storage is capable of reading millions of quantum bits in parallel, hence the verification of the quantum data medium can be attained much more effectively (Chuang et al., 2008).

The presented quantum data medium could be implemented in practice by a transmission-reflective type holographic optical disk and the readout process could be carried out by a holographic reader device. The rotation process uses a quantum CPU, implemented inside the reader device.

**Figure 18**   An experimental realization of the quantum copy-protection system

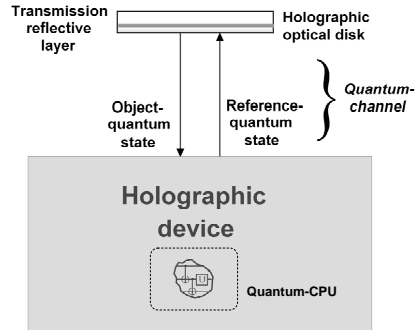

In the proposed model, the $\sigma_X, \sigma_Y$ and $\sigma_Z$ Pauli-transformations could be realized by holographic and optical elements. With the help of these optical elements, an arbitrary unitary quantum transformation could be prepared effectively (Chuang et al., 2008).

*5.1. The holographic optical data medium*

The quantum protected data medium is an information carrier that includes information in the form of a hologram. In order to increase protection against copying, the implemented hologram is the combination of a *transmission hologram* and a *reflection hologram*. The data and the key states on the quantum protected medium are encoded holographically. The object state is equal to the encoded *data or key* quantum state, while the reference state is equivalent to its orthogonal state.

**Figure 19** The reflective transmission holographic optical disk

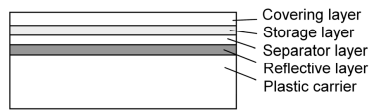

Such information carriers frequently serve as security seals for characterizing trademark products. The layer structure of reflective transmission holographic optical disk is illustrated in Figure 19.

# 6 Conclusions and future work

This paper introduces a fundamentally new method for protecting classical data by quantum states. The software Issuer stores the classical data in quantum states on the original distributed quantum-based data medium. The classical data on the data medium can be read out easily by anyone, but these classical bits are encoded bits. The decoding key rotates the angles of the quantum bits, thus some qubit values remain unchanged while some quantum states rotate to reverse their value and therefore the projected values of the classical bits. The absolute security of the proposed method rests on the laws of quantum dynamics. The operations can be implemented with some associated error, which decreases exponentially with the number of quantum states of the angle state. In further analysis of the proposed quantum-based data protection system, we would like to make an in-depth study of the applicability of quantum cloning machines, which can be used to produce the clones of quantum states.